\documentclass[twocolumn,prl,showpacs,aps,floatfix]{revtex4}
\usepackage{amssymb}
\usepackage{bm}
\usepackage{graphicx}
\usepackage[latin1]{inputenc}

\begin{document}
\title{State reconstruction of finite dimensional compound systems via local projective measurements and one-way classical communication}
\author{F. E. S. Steinhoff}
\email{steinhof@ifi.unicamp.br}
\author{M. C. de Oliveira}
\email{marcos@ifi.unicamp.br}
\affiliation{ Instituto de F\'\i
sica ``Gleb Wataghin'', Universidade Estadual de Campinas,
13083-970, Campinas, SP, Brazil}

\begin{abstract}
For a finite dimensional discrete bipartite system we find the relation between local projections performed by Alice and Bob's post-selected state dependence on the global state submatrices. With this result the joint state reconstruction problem for a bipartite system can be solved with strict local projections and one-way classical communication. The generalization to multipartite systems is straightforward. 
\end{abstract}
\maketitle

 Quantum state reconstruction relies  on the ability to measure a complete set of observables and further manipulation of numerical data in a way to describe unambiguously a quantum system state. Several approaches have been given regarding the reconstruction of sole and joint quantum system states \cite{Paris04,JMO}. Remarkable experiments have been realized, employing these reconstruction methods (See e.g. Refs. \cite{Kwiat99,Hafner05}). For a discrete Hilbert space,  quantum state reconstruction  is achieved through the determination of the complete set of real parameters describing the state \cite{James01}. For a continuum Hilbert space it relies on homodyne tomographic reconstruction \cite{JMO}. Unfortunately full state reconstruction is highly demanding. For a multipartite system composed by $N$ qubits, for example, it is well known that $4^N-1$ different real parameters (the generalized Stokes parameters), and thus an irreducible number of $3^N$ different settings are required. Supposing a qubit encoding over light polarization, the reconstruction is achieved through $2\times 3^N$ different measurements, with $2N$ photodetectors, most of them requiring multiple coincidences.  Although it is certainly not possible to reduce the number of parameters to be determined it is important to investigate alternative reconstruction schemes reducing actual experimental limitations imposed by multiple coincidence measurements, and/or the number of detectors.

In this paper we propose a scheme to reconstruct the state of a bipartite system of arbitrary finite dimension using only strict local projections and one-way classical communication. This scheme follows naturally from the answer to a more general question: if Alice performs projective measurements on her state, how this affects the description of Bob´s state? To answer this question, we adopt an operational approach based on the submatrices of the global system: after Alice´s measurement, Bob´s state will have some dependence on the submatrices of the initial global system. We obtain then a general condition that an arbitrary set of projections must fulfill in order to determine completely the global state. We analyze well-known and experimentally accessible projections, benefiting   from previous derivations of single system state reconstruction based on number state projectors \cite{Vaccaro95,Pregnell02, Thew02}, as well as the reconstruction of bipartite Gaussian continuous variable states employing local measurements and classical communication \cite{Rigolin09}. Focusing primarily on the simpler and important case in which Alice´s subsystem is two-dimensional (a qubit), we analyze the projections arising from her measurements of Pauli matrices - which are implementable for photons, spin-1/2 particles and two-level atoms \cite{Kwiat99,Hafner05,James01} - obtaining thus an experimentally feasible protocol for state reconstruction. We then focus on arbitrary bipartite dimension systems, analyzing a set of projections \cite{Gale68,Vaccaro95} to derive a state reconstruction protocol. Finally, we show a straightforward generalization of our findings for multipartite systems. The detections performed will not be dependent on coincidence verifications, and the number of detectors employed can be reduced in some cases. This is crucial for the important case of multiqubit state or joint correlations reconstruction, and thus for multipartite entanglement detection \cite{Thiago06}.

An arbitrary density matrix of a bipartite finite dimensional system shared by two parts, Alice and Bob, is recasted  in terms of well-suited $d_B\times d_B$ submatrices  $A_{ij}$ in the computational basis $\{|0,0\rangle, |0,1\rangle,\ldots, |0,d_B-1\rangle,|1,0\rangle,\ldots,|1,d_B-1\rangle,\ldots,|d_A-1,d_B-1\rangle\}$ (with $d_A$ and $d_B$ being the dimensions of Alice and Bob´s subsystems, respectively) as
\begin{eqnarray}
\rho = \left(\begin{array}{c c c}{A_{00}}&{\ldots}&{A_{0,d_A-1}}\\{\vdots}&
{\ddots}&{\vdots}\\{A_{0,d_A-1}^{\dagger}}&{\ldots}&{A_{d_A-1,d_A-1}}\end{array}\right).\label{dm}
\end{eqnarray}
The ordering of the computational basis is made in such a way as to simplify algebraic manipulations. We now derive two important results. 

\textit{Lemma 1}: 
Alice´s density matrix is given by
\begin{eqnarray}
\rho_A=Tr_B\rho = \left(\begin{array}{c c c}{TrA_{00}}&{\ldots}&{TrA_{0,d_A-1}}\\{\vdots}&
{\ddots}&{\vdots}\\{TrA_{0,d_A-1}^*}&{\ldots}&{TrA_{d_A-1,d_A-1}}\end{array}\right)
\end{eqnarray}
i.e., Alice´s density matrix element $(\rho_A)_{ij}$ is the trace of global state submatrix $A_{ij}$.

\textit{Proof.}  In the computational basis, $\rho = \sum_{i,k=0}^{d_A-1}\sum_{j,l=0}^{d_B-1}\rho_{ijkl}|ij\rangle \langle kl|$. Alice´s state is given by the partial trace $\sum_{\nu =0}^{d_B-1}{}_B\langle\nu |\rho |\nu\rangle_B$. An arbitrary term of this sum, ${}_B\langle\nu |\rho |\nu\rangle_B$, is easily seen to be $\sum_{i,k=0}^{d_A-1}\rho_{i\nu k\nu}|i\rangle\langle k|$ and Alice´s state is then $\rho_A=\sum_{\nu =0}^{d_B-1} {}_B\langle\nu |\rho |\nu\rangle_B = \sum_{\nu=0}^{d_B-1}\sum_{i,k=0}^{d_A-1}\rho_{i\nu k\nu}|i\rangle\langle k|$. An arbitrary element of $\rho_A$ is $(\rho_A)_{{\mu\eta}}=\langle\mu |\rho_A |\eta\rangle = \sum_{\nu =0}^{d_B-1}\rho_{\mu\nu\eta\nu}$ and an arbitrary element of the submatrix $A_{\mu\eta}$ is $(A_{\mu\eta})_{\alpha\beta}=\rho_{\mu\alpha\eta\beta}$. Finally, the trace is simply $TrA_{\mu\eta}=\sum_{\nu=0}^{d_B-1}\rho_{\mu\nu\eta\nu}=(\rho_A)_{\mu\eta}$. \hfill\rule{2mm}{2mm}

\textit{Proposition 1}:
 Let $|\psi\rangle = 
 \sum_{m=0}^{d_A-1}\alpha_m |m\rangle$ be a state in Alice´s subsystem and let $P_{|\psi\rangle}=|\psi\rangle\langle\psi |$ be the projector in this state. Then we have that Bob´s state after Alice´s projection in state $|\psi\rangle$ is given by
\begin{eqnarray}
\rho_B^{|\psi\rangle} = \frac{\sum_{m,n}\alpha_m\alpha_n^*A_{nm}}{\sum_{m,n}\alpha_m\alpha_n^*(\rho_A)_{nm}}. \label{theo}
\end{eqnarray}

\textit{Proof.}  Take $P_{|\psi\rangle} = \sum_{m,n=0}^{d_A-1}\alpha_m\alpha_n^*|m\rangle\langle n|$. Then it is easy to check that $(P_{\psi}\otimes I_B)\rho= \sum_{m,n,k=0}^{d_A-1}\sum_{j,l=0}^{d_B-1}\alpha_m\alpha_n^*\rho_{njkl}|mj\rangle\langle kl|$. Now, tracing out Alice´s subsystem we get $Tr_A[(P_{\psi}\otimes I_B)\rho]= \sum_{\nu =0}^{d_A-1}{}_A\langle\nu |\left(\sum_{m,n,k=0}^{d_A-1}\sum_{j,l=0}^{d_B-1}\alpha_m\alpha_n^*\rho_{njkl}|mj\rangle\langle kl|\right)|\nu\rangle_A = \sum_{m,n=0}^{d_A-1}\sum_{j,l=0}^{d_B-1}\alpha_m\alpha_n^*\rho_{njml}|j\rangle\langle l|$. We see then that in Bob´s subsystem $Tr_A[(P_{|\psi\rangle}\otimes I_B)\rho]$ is the same matrix as $\sum_{m,n=0}^{d_A-1}\alpha_m\alpha_n^*A_{nm}$. We conclude that Bob´s state after a projection in Alice´s subsystem is given by
\begin{eqnarray}
\rho_B^{|\psi\rangle} &=& \frac{Tr_A[(P_{|\psi\rangle}\otimes I_B)\rho]}{Tr[(P_{|\psi\rangle}\otimes I_B)\rho]}=\frac{\sum_{m,n}\alpha_m\alpha_n^*A_{nm}}{\sum_{m,n}\alpha_m\alpha_n^*TrA_{nm}},
\end{eqnarray}
and with the help of Lemma 1 we obtain  Eq. (\ref{theo}), \hfill\rule{2mm}{2mm}

\textit{(i) General reconstruction projections.} We give now the general procedures for reconstruction of bipartite states. Note that the expression in Proposition 1 can be rewritten as
\begin{eqnarray}
\sum_{m,n}\alpha_m\alpha_n^*A_{nm} = \rho_B^{|\psi\rangle}\left(\sum_{m,n}\alpha_m\alpha_n^*(\rho_A)_{nm}\right).
\end{eqnarray}
In general, the bipartite state reconstruction problem can be solved if we use a series of projections on different states $|\psi^{(\nu)}\rangle=\sum_i \alpha_i^{(\nu)}|i\rangle$, where $\nu=1,2,\ldots$; making these projections amounts to constructing the following system of equations:
\begin{eqnarray}
\sum_{m,n}\alpha_m^{(\nu)}\alpha_n^{(\nu)*}A_{nm} = \rho_B^{|\psi^{(\nu)}\rangle}\left(\sum_{m,n}\alpha_m^{(\nu)}\alpha_n^{(\nu) *}(\rho_A)_{nm}\right). \label{result}
\end{eqnarray}
Performing  suitable projections, so that we can invert the above system of equations, allows one to obtain the submatrices $A_{mn}$, and this is the same as to determine the state. So the condition to be fulfilled by \textit{any} set of projectors used by Alice is that the above system of equations must be invertible. An additional requirement is that Bob, or Alice, be able to perform local tomography of its state. General methods for state estimation such as maximal likelihood are standardly applied in imperfect tomography and are inherent to local reconstruction methods \cite{Paris04,JMO}. Thus to simplify we assume that Bob is able to perform local tomography with $N_B$ copies. 
\begin{figure}[!ht]
\includegraphics[width=8cm]{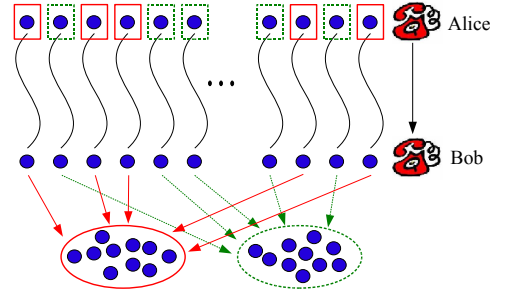}\vspace{-6cm}\caption{\label{fig-groups}(Color online) Sketch of the protocol for local reconstruction of a qubit-qudit joint state. \textit{(i)} Alice measures $\sigma_z$ projecting on $|0\rangle$ or $|1\rangle$ depending on the outcomes $\pm1$. \textit{(ii)} Alice measures $\sigma_x$ and $\sigma_y$ projecting in $|0\rangle\pm |1\rangle$, and $|0\rangle\pm i|1\rangle$, respectively. Alice results are classically communicated to Bob.}
\end{figure}

\textit{(ii) 1 Qubit $\otimes$ 1 qudit.}  Let us consider first the simplest case when Alice´s state is a qubit and Bob´s is an arbitrary one (qudit). This is an important example since in this case the reconstruction resembles the standard procedure for state reconstruction of qubit systems \cite{James01}, apart from the unnecessary coincidence detections. Here these are replaced by a communication protocol over a classical channel. 
  An arbitrary density matrix in the basis $\{|0,0\rangle,\ldots, |0,d_B-1\rangle,|1,0\rangle,\ldots,|1,d_B-1\rangle\}$ is given by $\rho =  \left(\begin{array}{c c}{A_{00}}&{A_{01}}\\{A_{01}^{\dagger}}&
{A_{11}}\end{array}\right)$, where $A_{ij}$ are $d_B\times d_B$ submatrices. Also, Alice´s density matrix can be written 
$\rho_A =\frac{1}{2} \left(\begin{array}{c c}{1+z_A}&{x_A-iy_A}\\{x_A+iy_A}&
{1-z_A}\end{array}\right)$ in terms of the Stokes parameters $1$, $x_A$, $y_A$, and $z_A$.
Alice's Pauli matrices´ projective measurements amounts to performing the projections $P_{|0\rangle}$, $P_{|1\rangle}$, $P_{|0\rangle \pm|1\rangle}$ and $P_{|0\rangle \pm i|1\rangle}$. Equations (\ref{result}) in this case read 
\begin{eqnarray}
A_{00} = \frac{1+z_A}{2}\rho_B^{|0\rangle}, &&A_{11} = \frac{1-z_A}{2}\rho_B^{|1\rangle}, \label{z}\\
A_{00}+A_{11}\pm(A_{01}+A_{01}^{\dagger})&=&(1\pm x_A)\rho_B^{|0\rangle\pm |1\rangle}, \label{x}\\
A_{00}+A_{11}\pm i(A_{01}-A_{01}^{\dagger})&=&(1\pm y_A)\rho_B^{|0\rangle\pm i|1\rangle}. \label{y}
\end{eqnarray}
The above system of equations for the submatrices is easily seen to be overdetermined. This results in a redundancy, which can be eliminated dealing appropriately with the measurement results. Such an issue appears naturally in the description of the protocol, which we do now. Suppose Alice and Bob share many copies of the system whose state they want to reconstruct. Suppose also that Bob needs $N_B$ copies of an arbitrary state in his part to perform a good local tomography. Alice starts measuring $\sigma_z$ on her states, performing projections $P_{|0\rangle}$ and $P_{|1\rangle}$ according to whether outcomes $\pm 1$ occur. Alice communicates her outcomes to Bob, who separates his states in two different subsets, according to these outcomes (see Fig.1). Alice continues the measurements until both outcomes occurred at least $N_B$ times. Bob then by any way performs the local tomographies of $\rho_B^{|0\rangle}$ and $\rho_B^{|1\rangle}$. Note that Alice´s outcomes are enough for Bob to obtain the value $z_A$ \footnote{This value is given by $z_A=\langle\sigma_z\rangle=\frac{n_+-n_-}{n_++n_-}$, where $n_{\pm}$ is the number of $\pm 1$ outcomes, respectively.}; by (\ref{z}), Bob obtains the diagonal submatrices $A_{00}$ and $A_{11}$. A subtle aspect should be noted here. What if Alice measures $\sigma_z$ and obtains too many outcomes $+1$ - and hence  very few $(-1)$ - so that Bob cannot estimate properly $\rho_B^{z-}$? In this case, the mean value $z_A=\langle\sigma_z\rangle$ is very close to $+1$: we can assume then that $A_{11}=\frac{1-z_A}{2}\rho_B^{z-}\approx 0_{d_B\times d_B}$, the null matrix of dimension $d_B\times d_B$. 

The procedure to obtain the off-diagonal submatrix $A_{01}$ has a slight modification. Alice measures $\sigma_x$ on her states, obtaining outcomes $\pm 1$, what amounts to respective projections $P_{|0\rangle\pm |1\rangle}$. Alice communicates her outcomes to Bob, who separates his states in two different subsets, according to the outcomes. The modification here is that Alice will measure \textit{until} one of the outcomes occurred $N_B$ times. When this situation occurs, she can stop to measure; Bob then by any way determines the density matrix corresponding to this outcome. Using (\ref{x}) and the already determined diagonal submatrices $A_{00}$, $A_{11}$ Bob obtains $A_{01}+A_{01}^{\dagger}$. Once again Alice´s outcomes are enough for Bob to obtain the value $x_A$; note also that he needs to use only one of the equations (\ref{x}), eliminating half of the redundancy. 
Repeating this procedure for $\sigma_y$ it should be clear that Bob obtains $A_{01}-A_{01}^{\dagger}$ by (\ref{y}) and hence $A_{01}$; the other half of the redundancy is eliminated in this step. Alice and Bob then reconstruct the state using only strict local projective measurements and one-way classical communication. It is important to realize that the parameters $x_A$, $y_A$ and $z_A$ will be very well estimated. For a two-qubit polarization state implemented in optical systems, $A_{ij}$ are $2\times2$ submatrices, which can be determined by Bob through the same kind of measurements employed by Alice. The number of different experimental settings is $3^2=9$, as in the usual state reconstruction \cite{James01}, but detections do not need to be in coincidence. As a consequence, that will require only $2$ photodetectors as opposed to the $4$ used in the usual state tomography \cite{James01}. The only requirement for Bob are the results $\pm 1$ of Alice´s measurements, so that he can discriminate his states.

\textit{(ii) 1 Qudit $\otimes$ 1 qudit.} We present now a set of projectors suited to the case when both subsystems are of arbitrary dimensions. These projectors may be implemented in optical systems as demonstrated in \cite{Pregnell02,Thew02} or employing Stern-Gerlach apparatuses \cite{Gale68}. Most importantly, they suffice to determine the elements of Alice´s density matrix, which are necessary in the protocol. 
These projectors are $P_{|j\rangle}$ (with $j$ varying from $0$ to $d_A -1$), $P_{|j\rangle \pm |k\rangle}$ and $P_{|j\rangle\pm i|k\rangle}$ (with $j<k$ and both varying from $0$ to $d_A-1$). 
 Explicitly Eq. (\ref{result}) for these projectors writes as
\begin{eqnarray}
A_{jj} &=& \rho_B^{|j\rangle}(\rho_A)_{jj}, \label{x1}\\
\Delta_{jk} &=&  \left\{(\rho_A)_{jj} + (\rho_A)_{kk}\pm 2Re[(\rho_A)_{jk}]\right\}\rho_B^{|j\rangle\pm |k\rangle}, \label{x2}\\
\Omega_{jk}&=& \left\{(\rho_A)_{jj} + (\rho_A)_{kk} \mp 2Im[(\rho_A)_{jk}]\right\}\rho_B^{|j\rangle\pm i|k\rangle},\label{x3}
\end{eqnarray}
where $\Delta_{jk}\equiv A_{jj}+A_{kk}\pm(A_{jk}+A_{jk}^{\dagger})$, $\Omega_{jk} \equiv A_{jj}+A_{kk}\pm i(A_{jk}-A_{jk}^{\dagger})$, with $k>j=0,\cdots,d_A-1$.
The resulting system of equations for the submatrices is once again overdetermined. We explain now briefly the protocol, which resembles the qubit-qudit protocol in many steps.

Suppose that Alice and Bob share many copies of the state they want to reconstruct and that Bob needs $N_B$ copies of an arbitrary state in his subsystem to perform a good local tomography. To determine a diagonal submatrix $A_{jj}$, Alice starts measuring $P_{|j\rangle}$ on her states, communicating to Bob whether or not (outcomes $1$ or $0$, respectively) the projection actually happened. She repeats this procedure until there happened $N_B$ outcomes $1$. Bob then determines by any way the density matrix of his states corresponding to outcomes $1$. As Alice´s element $(\rho_A)_{jj}$ is precisely the mean value $\langle P_{|j\rangle}\rangle$, Bob can get this value with Alice´s outcomes and hence, by (\ref{x1}), he determines $A_{jj}$. However, if $(\rho_A)_{jj}$ is too small, then we can approximate $A_{jj}$ by the null matrix, as done in the qubit-qudit case for the diagonal submatrices. So, we can reconstruct all diagonal submatrices.

To determine a off-diagonal submatrix $A_{jk}$ ($j\neq k$), Alice chooses first to perform one of the projections $P_{|j\rangle\pm |k\rangle}$ on her states. Due to the redundancy in equations (\ref{x2}), only one of them is enough to determine $A_{jk}+A_{jk}^{\dagger}$. It is important, however, that Alice be able to perform the other projection. Alice then perform the projection chosen on her states, communicating to Bob whether or not (outcomes $1$ or $0$, respectively) the projection actually happened. She repeats this procedure until there happened $N_B$ outcomes $1$. Bob then determines by any way the density matrix of his states corresponding to outcomes $1$. Alice´s element $Re[(\rho_A)_{jk}]$ is determined if one observes the property $P_{|j\rangle + |k\rangle}+P_{|j\rangle - |k\rangle}=2(P_{|j\rangle}+P_{|k\rangle})$. Then, we have that $\langle P_{|j\rangle + |k\rangle}\rangle+\langle P_{|j\rangle - |k\rangle}\rangle=2(\rho_A)_{jj}+2(\rho_A)_{kk}=g_{jk}$, where $g_{jk}$ is a constant value which is already determined, since Alice´s diagonal elements are known. As $Re[(\rho_A)_{jk}]=\langle P_{|j\rangle + |k\rangle}\rangle - \langle P_{|j\rangle - |k\rangle}\rangle$ \cite{Gale68,Vaccaro95}, we can take $Re[(\rho_A)_{jk}]=2\langle P_{|j\rangle + |k\rangle}\rangle -g_{jk}$ or $Re[(\rho_A)_{jk}]=g_{jk}-2\langle P_{|j\rangle - |k\rangle}\rangle$. This idea can also be used in the situation where the projector chosen has a too small mean value, impossibiliting Bob to have $N_B$ copies on the corresponding subset. We can shift to the other projector, which will have a compensating greater value. There is, however, a third extreme situation,  when both mean values are small. But, in this case,  $(\rho_A)_{jj}$ and $(\rho_A)_{kk}$ are necessarily small as well, and the value $(\rho_A)_{jj} + (\rho_A)_{kk} \pm 2Re[(\rho_A)_{jk}]$ in (\ref{x2}) will be too small; then we can approximate $A_{jk}+A_{jk}^{\dagger}$ by the null matrix. In all situations, by (\ref{x2}) Bob determines $A_{jk}+A_{jk}^{\dagger}$. 

Finally, Alice chooses to perform one of the projections $P_{|j\rangle\pm i|k\rangle}$ and repeats the above procedure. The property $P_{|j\rangle + i|k\rangle}+P_{|j\rangle - i|k\rangle}=P_{|j\rangle}+P_{|k\rangle}$ will allows Bob to get $Im[(\rho_A)_{jk}]$. Bob then finds $A_{jk}-A_{jk}^{\dagger}$ by (\ref{x3}) and hence determines $A_{jk}$. As all submatrices were determined, the state is reconstructed, using only local projections and one-way classical communication. In this sense, coincidence requirements are unnecessary in tomography of compound systems. Also, as Alice and Bob acts locally on their respective systems, in some cases \footnote{This will depend on the experimental implementation of the projections. For example, if one uses Stern-Gerlach apparatus to generate them \cite{Gale68}, the diagonal projectors $P_{|j\rangle}$ will be generated by the observable $J_z$. Obviously one can then eliminate one detector when measuring the diagonal elements, due to the normalization condition on the state. However, the off-diagonal projectors are not all generated at the same time and thus we cannot make such an elimination.} Alice can eliminate $1$ detector in her subsystem, due to the completeness relations $\sum_{j=0}^{d_A-1} P_{|j\rangle}=I$, $\sum_{j,k=0}^{d_A-1}P_{|j\rangle + |k\rangle}+P_{|j\rangle - |k\rangle}=(d_A-1)I$, $\sum_{j,k=0}^{d_A-1}P_{|j\rangle + i|k\rangle}+P_{|j\rangle - i|k\rangle}=(d_A-1)I$ - where we assume $j<k$. If Bob uses the same set of projectors, he can also eliminate $1$ detector by the same reason; we get thus an economy of $2$ detectors. These considerations are even more dramatic when considering multipartite settings.

\textit{(iii) Multiple Qudits.}  When dealing with a system composed of $N$ subsystems, the state space is composed of the tensor product of the individual Hilbert spaces: $H=H_1\otimes H_2\otimes\ldots\otimes H_N$. We can view this system as a bipartite one, considering the tensor product of one of them with the others\begin{equation} H = \underbrace{H_1}_{H_{A^{(1)}}}\otimes\underbrace{H_2\otimes\ldots\otimes H_N}_{H_{B^{(1)}}}=H_{A^{(1)}}\otimes H_{B^{(1)}}.\end{equation} We can then apply the protocol for a bipartite state reconstruction to this bipartite system, making the projections on $H_1$ and sending outcomes to $H_2$. After the projection, $H_1$ will be left in a pure state and so will not affect subsequent operations made on $H_{B^{(1)}}$. Tomography of the various states $\rho_{B^{(1)}}^{|j\rangle}$, $\rho_{B^{(1)}}^{|j\rangle\pm |k\rangle}$, $\rho_{B^{(1)}}^{|j\rangle\pm i|k\rangle}$ must be realized in order to perform the task of reconstruction (if one uses the projections $P_{|i\rangle}$, $P_{|j\rangle \pm|k\rangle}$, $P_{|j\rangle \pm i|k\rangle}$). This is achieved by repeating the above procedure \begin{equation} H_{B^{(1)}}= \underbrace{H_2}_{H_{A^{(2)}}}\otimes\underbrace{H_3\otimes\ldots\otimes H_N}_{H_{B^{(2)}}}=H_{A^{(2)}}\otimes H_{B^{(2)}}. \end{equation} After $N-1$ repetitions of this procedure we can reconstruct the global state density matrix. Coincidence is an unnecessary requirement, with an economy of $N$ detectors for the set of projectors considered. For the case of $n$ qubits, one concludes that reconstruction task requires $n$ detectors, in contrast to usual $2n$ employed, without any need of coincidence requirements.

\textit{Conclusions.} Quantum state reconstruction is a extremely important task in any implementation of quantum computation or quantum communication protocol. It is supposed to determine unambiguously a single or joint quantum system state. Unfortunately its implementation for composed systems is severely compromised for the requirement of joint measurements, which increases in number for both increasing the individual Hilbert space dimension and/or number of subsystems. We have given an alternative procedure for reconstruction based on local measurements and classical communication that enables bipartite or multipartite quantum systems of arbitrary Hilbert space dimension to be reconstructed without the need of joint or coincidence measurements. Beyond the natural importance in quantum computation and quantum information, we believe it has important implications for the reconstruction of manyparticle system state from simple measurements of local moments, such as magnetization for spin systems \cite{Thiago06}. 
 The actual reconstruction of such a kind of system state through this method will be thus a remarkable achievement.

This work is supported by CNPq and FAPESP through the National Institute for Quantum Information.

\end{document}